\documentclass[traditabstract]{aa}
\usepackage{epsfig}
\usepackage{natbib}
\usepackage{graphics}
\usepackage{color} 
\usepackage{ulem}
\usepackage{times}
\usepackage{color}
\newfont{\tss}{cmssdc10 scaled 950}

\begin{document}

\title{Star-forming galaxies with hot dust emission 
in the Sloan Digital Sky Survey discovered by the 
Wide-field Infrared Survey Explorer (WISE)}


\author{Y. I.\ Izotov \inst{1,2,3}
\and N. G.\ Guseva \inst{1,2}
\and K. J.\ Fricke \inst{1,4}
\and C.\ Henkel \inst{1,5}
}
\offprints{Y.I. Izotov, izotov@mao.kiev.ua}
\institute{          Max-Planck-Institut f\"ur Radioastronomie, 
                     Auf dem H\"ugel 
                     69, 53121 Bonn, Germany
\and
                     Main Astronomical Observatory,
                     Ukrainian National Academy of Sciences,
                     Zabolotnoho 27, Kyiv 03680,  Ukraine
\and
                     LUTH, Observatoire de Paris, CNRS, 
                     Universite Paris Diderot,
                     Place Jules Janssen 92190 Meudon, France
\and
                     Institut f\"ur Astrophysik, 
                     G\"ottingen Universit\"at, Friedrich-Hund-Platz 1, 
                     37077 G\"ottingen, Germany
\and
                     Astronomy Department, King Abdulaziz University, 
                     P.O. Box 80203, Jeddah, Saudi Arabia
}

\date{Received \hskip 2cm; Accepted}
\abstract{
We present the results of a search for 
Sloan Digital Sky Survey (SDSS) emission-line
galaxies with very red 3.4$\mu$m - 4.6$\mu$m $(W1-W2)$ colours in the 
{{\sl Wide-field Infrared Survey Explorer}} ({{\sl WISE}}) Preliminary Release 
Source Catalogue (PRSC) aiming to find objects with hot dust emission.
For this purpose we considered a sample of $\sim$ 16000 galaxies 
with strong emission lines
selected out of a total of $\sim$ 900000 SDSS spectra
and identified them with the PRSC sources.
We find that $\sim$ 5000 sources out of the $\sim$ 16000 SDSS galaxies are 
present in the PRSC, including $\sim$ 1000 galaxies with sufficiently strong 
[O {\sc iii}] $\lambda$4363 emission lines to permit 
reliable determinations of the oxygen abundance. No correlation of $W1 - W2$ 
with metallicity is found. On the other hand, there is clear
evidence for a redder $W1 - W2$ index in galaxies with higher H$\beta$
luminosity and higher H$\beta$ equivalent width, implying that strong UV 
radiation from young starbursts efficiently heats interstellar dust to 
high temperatures. However, galaxies with very 
red colours $W1 - W2$ $>$ 2 mag, 
similar to that in the local extreme star-forming galaxy SBS 0335--052E, are 
very rare. In addition to three previously 
known sources, which are not present in our sample, we found only four such 
galaxies.}
\titlerunning{New star-forming galaxies with hot dust emission}
\authorrunning{Y.I.Izotov et al.}
\maketitle


\section{Introduction}

Nearby star-forming emission-line galaxies play an important role for our
understanding of star-formation processes in low-metallicity environments, and 
they can be considered as local counterparts or ``analogs'' of high-redshift 
star-forming Lyman-break galaxies (LBGs). Recently, \citet{H05} 
identified nearby ($z$ $<$ 0.3) ultraviolet-luminous galaxies (UVLGs) selected
from the {\sl Galaxy Evolution Explorer} ({\sl GALEX}). Eventually, compact
UVLGs were called Lyman-break analogs (LBAs). They resemble LBGs in several 
respects.  In particular, their metallicities are 
subsolar, and their star-formation rates ($SFR$s) of 
$\sim$ 4 -- 25 $M_\odot$ yr$^{-1}$ are overlapping with those for LBGs.
Recently, \citet{C09} selected a sample of 251 compact strongly 
star-forming galaxies
at $z$ $\sim$ 0.112 -- 0.36 on the basis of their intense green colour on the 
Sloan Digital Sky Survey (SDSS) images (``green pea'' galaxies), which again 
are similar to LBGs owing to their low metallicity and high $SFR$s. 
\citet{I11} extracted a sample of 803 star-forming luminous compact
galaxies (LCGs) with hydrogen H$\beta$ luminosities 
$L$(H$\beta$) $\geq$ 3$\times$10$^{40}$ erg s$^{-1}$ and H$\beta$ equivalent
widths EW(H$\beta$) $\geq$ 50\AA\ from SDSS spectroscopic data.
These galaxies have properties 
similar to ``green pea'' galaxies but are distributed over a wider
range of redshifts $z$ $\sim$ 0.02 - 0.63. The $SFR$s of LCGs are high 
$\sim$ 0.7 -- 60 $M_\odot$ yr$^{-1}$ and overlap with those of LBGs.
\citet[see also \citet{G09}]{I11} showed that LBGs, LCGs, luminous metal-poor
star-forming galaxies \citep{Hoyos05}, extremely metal-poor 
emission-line galaxies at $z$ $<$ 1 \citep{Kakazu07}, 
and low-redshift blue compact
dwarf (BCD) galaxies with strong star-formation activity 
obey a common luminosity-metallicity relation. 
Therefore, it is important to study nearby star-forming galaxies over a wide 
range of luminosities and metallicities to shed light on physical conditions
and star-formation history in high-redshift galaxies 
even though most metal-deficient and 
low-luminosity high-redshift galaxies are still awaiting their detection.

\begin{figure*}[t]
\centering{
\hbox{\includegraphics[angle=-90,width=0.39\linewidth]{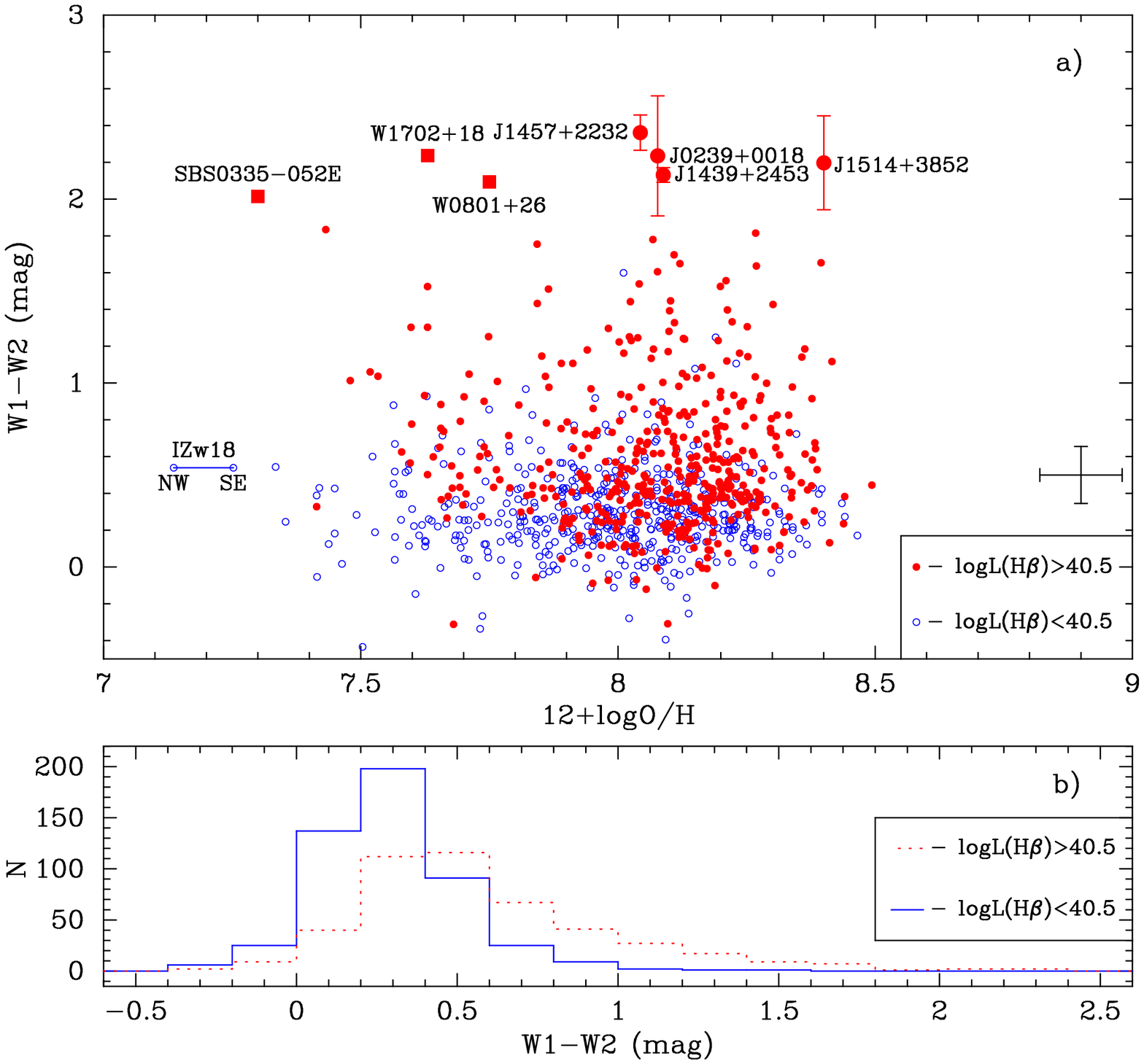} 
\hspace{0.3cm}\includegraphics[angle=-90,width=0.39\linewidth]{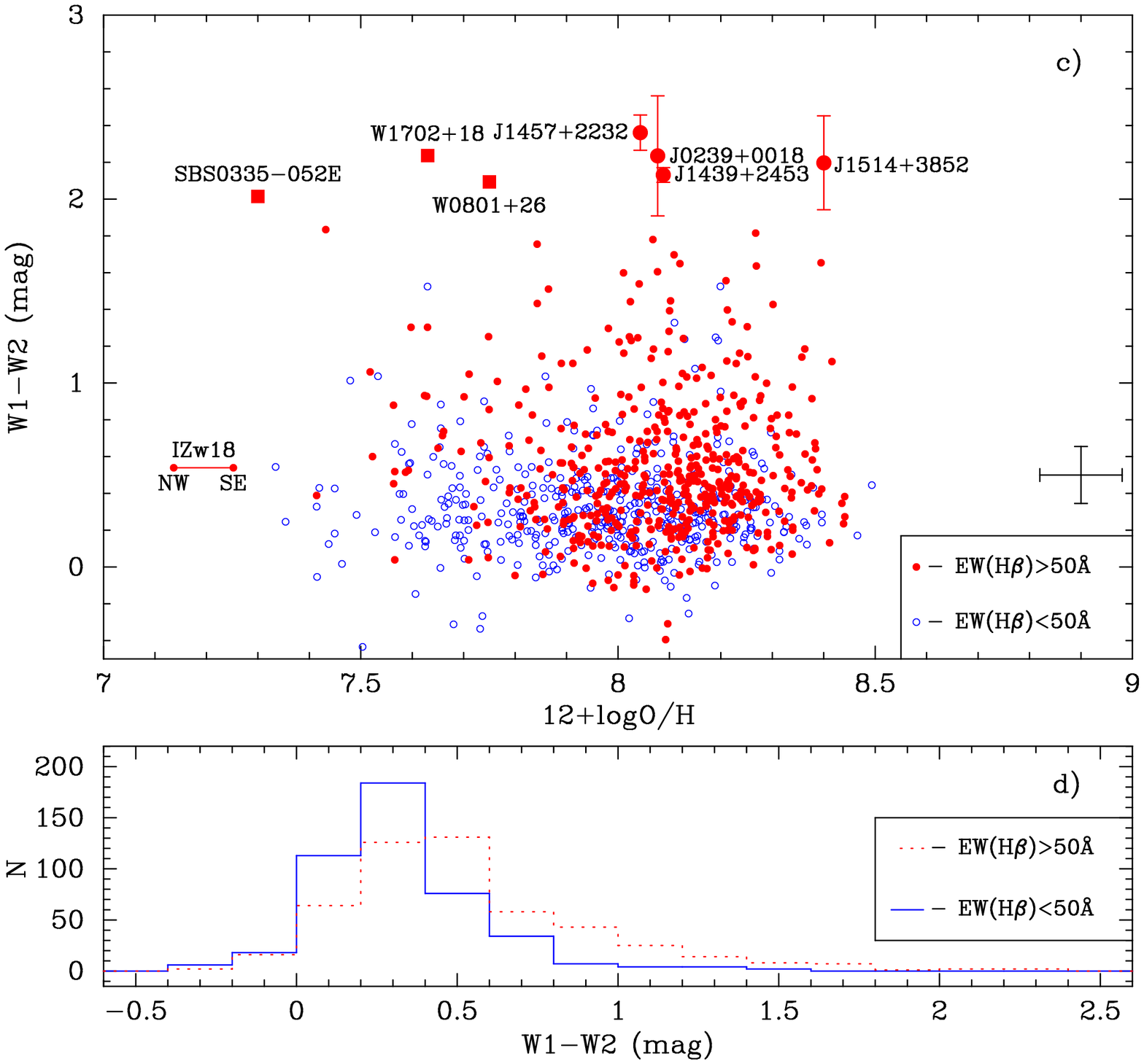}}
} 
\caption{{\bf (a)} Dependence of the $W1-W2$ colour on the oxygen abundance
12+logO/H for a sample of $\sim$ 1000 galaxies. Galaxies with 
H$\beta$ luminosity $L$(H$\beta$) $\geq$ 3$\times$10$^{40}$ erg s$^{-1}$,
corresponding to $SFR$(H$\beta$) $\geq$ 0.7 $M_\odot$ yr$^{-1}$, are shown
by red filled circles while galaxies with 
$L$(H$\beta$) $<$ 3$\times$10$^{40}$ erg s$^{-1}$ are shown by
blue open circles. The NW and SE components of the low-metallicity BCD
I Zw 18 are labelled and connected with a solid line.
Newly identified galaxies with $W1-W2$ $>$ 2 mag are shown
by large red filled circles, the three galaxies discussed by 
\citet{G11} are shown by large red filled squares. 
Typical error bars are shown in the lower right corner.
{\bf (b)} 
Histograms of $W1-W2$ 
distributions for galaxies with 
$L$(H$\beta$) $\geq$ 3$\times$10$^{40}$ erg s$^{-1}$ (red dotted line)
and with $L$(H$\beta$) $<$ 3$\times$10$^{40}$ erg s$^{-1}$
(blue solid line). {\bf (c)} and {\bf (d)} same as {\bf (a)} and {\bf (b)}, 
respectively, but the sample is split
into galaxies with EW(H$\beta$) $\geq$ 50\AA\ (red symbols and lines) and
EW(H$\beta$) $<$ 50\AA\ (blue symbols and lines).
}
\label{fig1}
\end{figure*}

The {\sl Infrared Space Observatory} ({\sl ISO}), {\sl Spitzer}, and most
recently the {\sl Wide-field Infrared Survey Explorer} 
\citep[{\sl WISE}, ][]{W10}
open up the opportunity to probe properties of star-forming galaxies in the
mid-infrared range (MIR) $\sim$ 3.5 -- 24 $\mu$m, the range of warm and hot 
dust. The {\sl WISE} mission has an advantage because it is 
directed to produce an all-sky photometric survey at wavelengths 
3.4$\mu$m ($W1$),
4.6$\mu$m ($W2$), 12$\mu$m ($W3$) and 22$\mu$m ($W4$) with a sensitivity at 
$\sim$ 12 - 24 $\mu$m that is $\sim$ 1000 times higher than that of the 
{\sl Infrared Astronomical Satellite} ({\sl IRAS}) and has an angular 
resolution of $\sim$ 6\arcsec\ at 3.4$\mu$m.

\citet{T99} first showed (from {\sl ISO} spectroscopy) that one of the 
most metal-deficient BCDs, SBS 0335--052E 
\citep[e.g. ][]{I90}, is extraordinarily bright in the
MIR range, implying a large amount of warm ($\sim$ 100 - 300K) dust. Later,
\citet{H04} based on {\sl Spitzer} spectra have confirmed the presence of
warm dust in SBS 0335--052E and found that the dust emission peaks at
a wavelength of $\sim$ 28 $\mu$m, much shorter than that for the bulk
of star-forming galaxies. 
Ground-based infrared spectroscopy of SBS 0335--052E
by \citet{H01} revealed that the continuum at shorter wavelengths in the range 
3.4 - 4 $\mu$m is rising in the direction of longer wavelengths. 
Later, \citet{G11} found that the {\sl WISE} 3.4 - 4.6 $\mu$m ($W1-W2$) colour
of SBS 0335--052E is very red, $>$ 2 mag. This longward rising of MIR emission
implies the presence of hot (up to 1000K) dust emission. 

The properties of SBS 0335--052E in the MIR range
are very different from that in another extremely metal-deficient BCD
with similar metallicity, I Zw 18. \citet{HH04} attributed these differences
to different modes of star formation, an ``active'' mode in very compact
regions of SBS 0335--052E and a ``passive'' mode in the more diffuse 
interstellar medium of I Zw 18.

\citet{G11} argued that {\sl WISE} can be an efficient tool in searching for 
other star-forming galaxies with appreciable hot dust emission and demonstrated
the truth of this statement by finding of three low-metallicity BCDs with 
$W1-W2$ $>$ 2 mag. However, they noted that these galaxies are rare.
In the present paper we attempt to find new star-forming galaxies selected
from the Data Release 7 (DR7) of the SDSS with red 
{\sl WISE} $W1-W2$ colours in the Preliminary Release 
Source Catalogue (PRSC), which covers 57\% of the sky.

\begin{figure*}
\centering{
\includegraphics[angle=-90,width=0.39\linewidth]{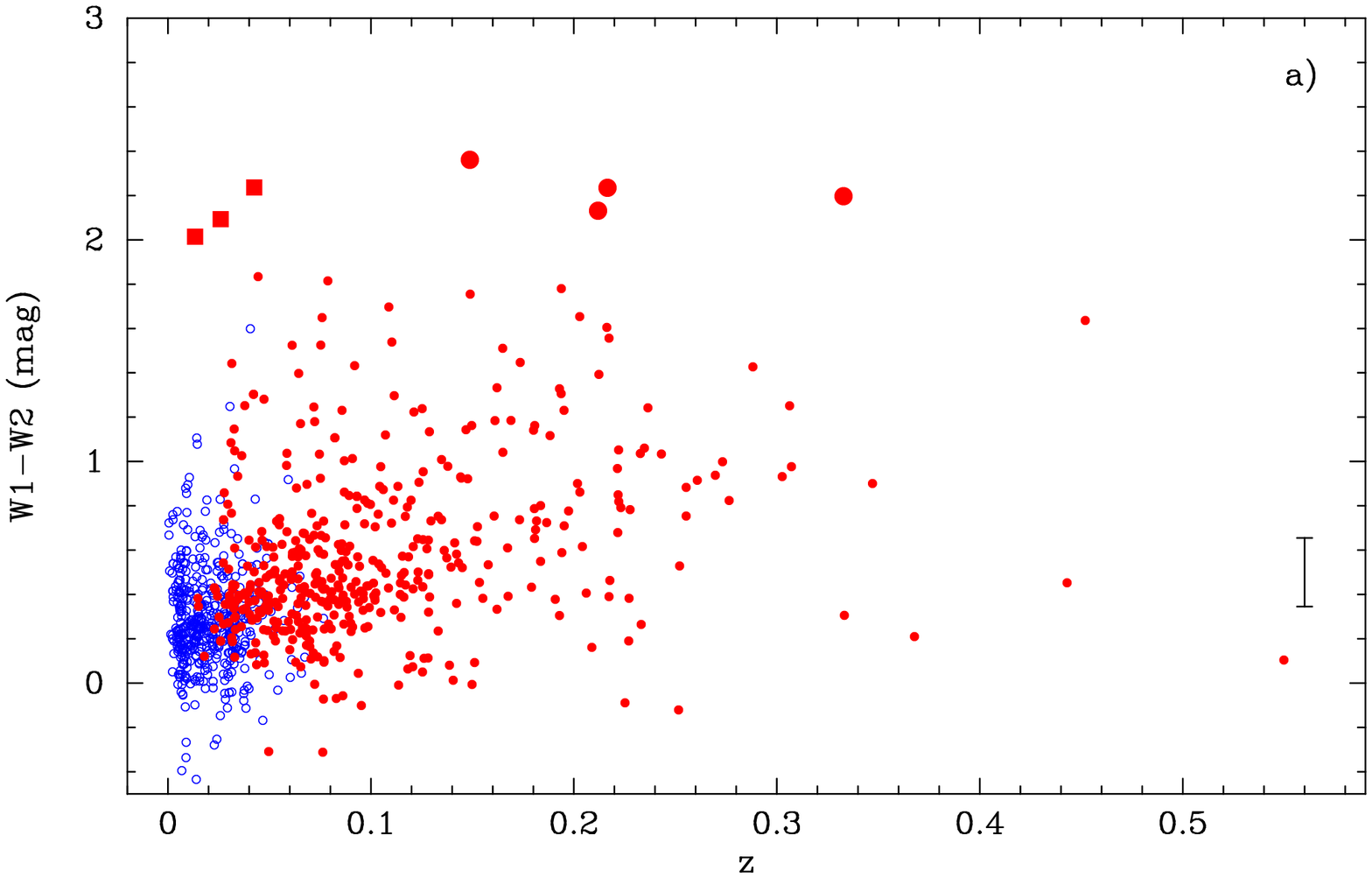}
\hspace{0.3cm}\includegraphics[angle=-90,width=0.39\linewidth]{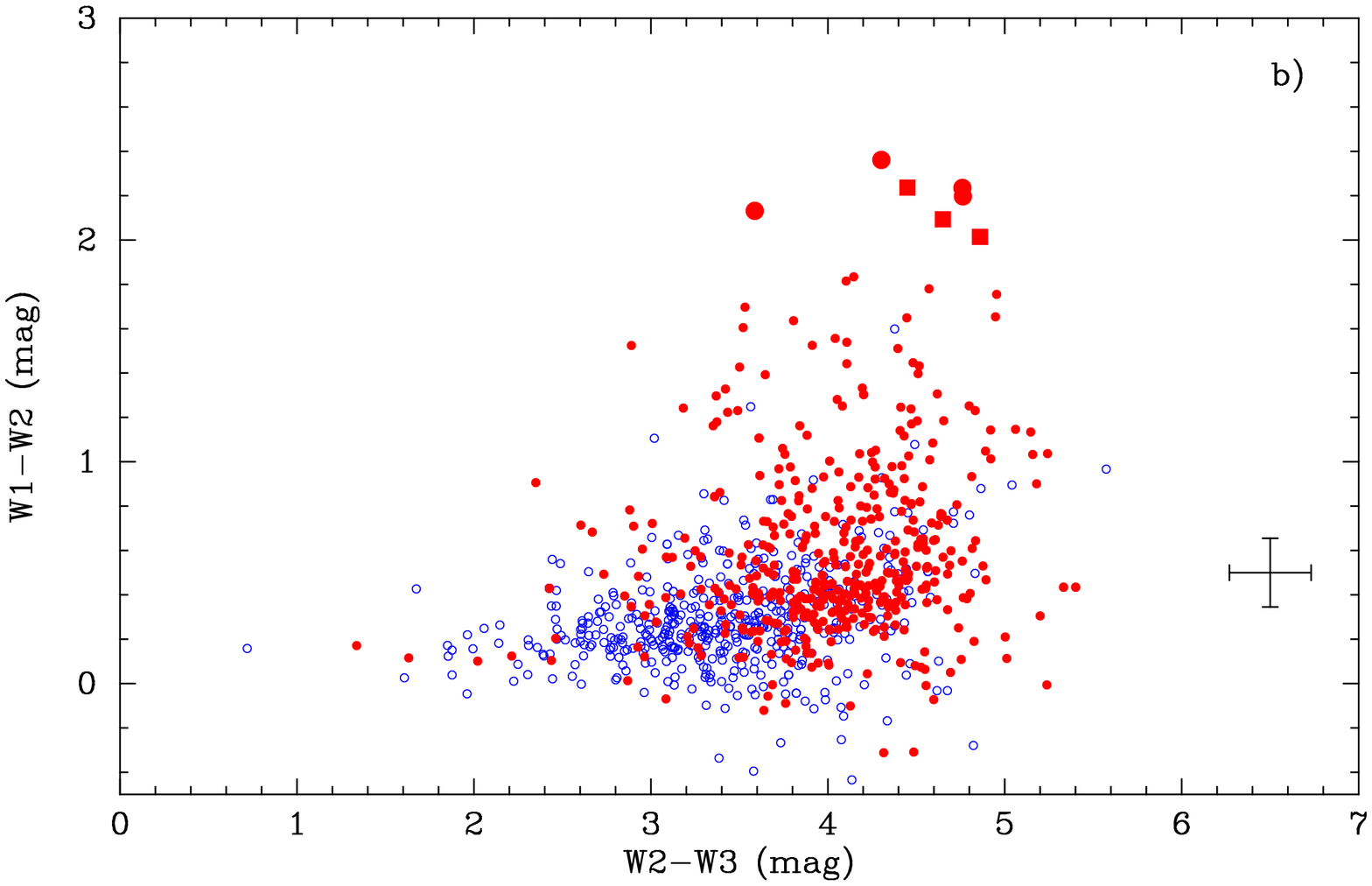}
} 
\caption{{\bf (a)} Dependence of $W1-W2$ colour on redshift $z$
for the sample of $\sim$ 1000 galaxies. {\bf (b)} $(W1-W2)$ vs. 
$(W2-W3)$ colour-colour diagram.
Symbols are the same as in Fig. \ref{fig1}a.}
\label{fig2}
\end{figure*}


\begin{table*}
\caption{Properties of newly identified star-forming galaxies with red $W1-W2$ colours \label{tab1}}
\begin{tabular}{lcccc} \hline\hline
Property                             &J0219+0018          &J1439+2453          &J1457+2232          &J1514+3852          \\ \hline
R.A.(J2000)                          & 02:39:00.79        & 14:39:05.24        & 14:57:35.14        & 15:14:08.63        \\
Dec.(J2000)                          &+00:18:35.88        &+24:53:53.39        &+22:32:01.79        &+38:52:07.31        \\
$z$                                  &0.2166              &0.2119              &0.1488              &0.3329              \\
SDSS $g$ (mag)                       &20.72               &19.54               &19.43               &20.44               \\
$W1$ (mag)                           &17.80$\pm$0.28      &14.26$\pm$0.03      &16.22$\pm$0.08      &17.60$\pm$0.23      \\
$W2$ (mag)                           &15.57$\pm$0.16      &12.13$\pm$0.02      &13.85$\pm$0.05      &15.40$\pm$0.12      \\
$W3$ (mag)                           &10.81$\pm$0.09      & 8.55$\pm$0.03      & 9.55$\pm$0.04      &10.64$\pm$0.07      \\
$W4$ (mag)                           & 7.65$\pm$0.12      & 6.16$\pm$0.05      & 6.54$\pm$0.06      & 7.77$\pm$0.13      \\
$L$(H$\beta$) (erg s$^{-1}$)         &2.8$\times$10$^{41}$&1.2$\times$10$^{42}$&4.8$\times$10$^{41}$&2.6$\times$10$^{41}$\\
$SFR$(H$\beta$) ($M_\odot$ yr$^{-1}$)&    6.1             &    26.5            &    10.6            &     5.8            \\
$M_*/M_\odot$$^a$                    &1.9$\times$10$^8$   &2.9$\times$10$^8$   &1.3$\times$10$^9$   &2.1$\times$10$^9$   \\
$M$(burst)/$M_\odot$$^a$             &3.0$\times$10$^7$   &9.9$\times$10$^7$   &4.7$\times$10$^7$   &4.8$\times$10$^7$   \\
EW(H$\beta$) (\AA)                   &   155              &    174             &    205             &    105             \\
$t$(burst) (Myr)$^a$                 &   3.9              &    3.6             &    3.1             &    3.9             \\
12+logO/H                            &     8.08           &     8.09           &     8.04           &     8.40           \\
\hline \\
\end{tabular}

$^a$Stellar masses $M_*$, starburst masses $M$(burst) and ages $t$(burst) are derived from SDSS spectra
according to prescriptions of \citet{I11}. 
\end{table*}


\section{Selection criteria \label{sel}}

   We used the whole spectroscopic data base of the SDSS DR7 \citep{A09},
which comprises $\sim$ 900000 spectra of galaxies to select 
a sample of $\sim$ 16000 spectra with strong emission lines, 
excluding those that show evidence
for AGN activity. Thus, our sample includes the majority
of SDSS star-forming galaxy spectra with EW(H$\beta$) $\geq$ 10\AA. 
A small fraction of
these spectra ($\sim$ 2\%) represents individual H {\sc ii} regions in nearby spirals but
the overwhelming majority is composed of integrated spectra of galaxies from 
farther distances.
The coordinates of SDSS selected galaxies are used to identify sources in 
the {\sl WISE} PRSC within a circular aperture of 10\arcsec\ in radius.
In total, $\sim$ 5000 {\sl WISE} sources were identified with SDSS galaxies
out of the sample of $\sim$ 16000 galaxies. 
Nearly 1000 galaxies out of $\sim$ 5000 galaxies detected by {\sl WISE} have 
sufficiently strong [O {\sc iii}] $\lambda$4363 emission lines in their SDSS spectra 
with a line flux error not exceeding 50\%, allowing for a reliable 
oxygen abundance determination. We analyse the MIR properties of this sample 
of $\sim$ 1000 galaxies in Section \ref{res}.

\section{Results \label{res}}

Several galaxy components can contribute to
the emission in the 3.4 -- 4.6 $\mu$m range: stars, ionised gas,
polycyclic aromatic hydrocarbon (PAH) emission, and hot
dust. Generally, the PAH emission is weak in a low-metallicity environment 
\citep{E08,H10}. Stellar (with an effective temperature $\geq$ 3000K)
and interstellar ionised gas emission is characterised by $W1-W2$ colours 
of $\sim$ 0 - 0.4 mag. In particular, the $W1-W2$ $\sim$ 0.5 mag of I Zw 18,
one of the most metal-deficient BCDs, is consistent with that of
stellar and ionised gas emission only.
On the other hand, a colour excess above $W1-W2$ $\sim$ 0.4 mag
could be an indication of hot dust with a temperature of several 
hundred Kelvin. In particular, $W1-W2$ colours 
of 3, 2, and 1 mags 
correspond to black body temperatures of 350, 500, and 900K, respectively. 
This colour excess also depends on the relative contribution of hot dust 
emission to the total emission. Therefore, while it clearly points at hot
dust in extreme cases, the $W1-W2$ colour alone is not 
sufficient for a precise determination of the dust temperature. 
The spectral energy 
distribution is needed in a wide wavelength range to fit and 
subtract stellar and gaseous emission.

\begin{figure}
\hbox{
\includegraphics[angle=0,width=0.75\linewidth]{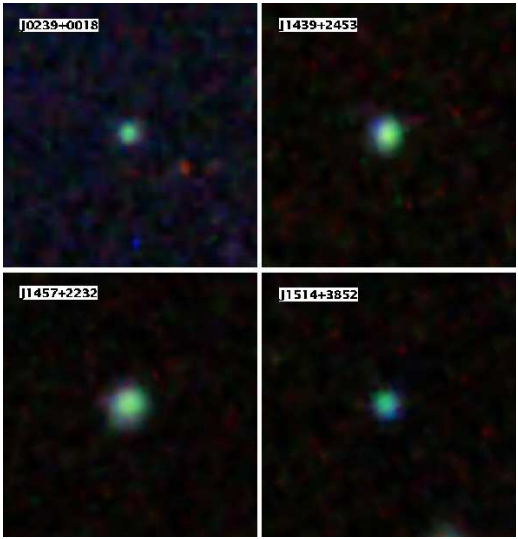} 
}
\caption{20\arcsec$\times$20\arcsec\ SDSS images of galaxies with
$W1-W2$ $>$ 2 mag.} 
\label{fig3}
\end{figure}

\begin{figure}
\centering{
\includegraphics[angle=-90,width=0.99\linewidth]{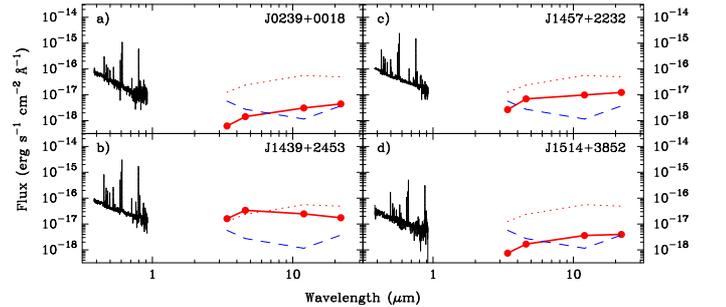}} 
\caption{
Observed spectral energy distributions (SED) of galaxies with 
$W1-W2$ $>$ 2 mag, which include SDSS optical spectra (black solid lines)
and {\sl WISE} MIR photometric data in all four bands (red symbols and solid lines).
For comparison in all panels are shown MIR SEDs for the ``active'' BCD 
SBS 0335--052E (red dotted lines) and for the ``passive'' BCD I Zw 18 
(blue dashed lines). 
}
\label{fig4}
\end{figure}

In Fig. \ref{fig1}a we show the dependence of $W1-W2$ on the oxygen
abundance 12+logO/H for the sample of $\sim$ 1000 galaxies with the best
determined oxygen abundances (see Sect. \ref{sel}). 
The sample is split into two 
subsamples of objects with the H$\beta$ luminosity 
$L$(H$\beta$) $\geq$ 3$\times$10$^{40}$ erg s$^{-1}$ (small red filled circles)
corresponding to a star-formation rate $SFR$ $\geq$ 0.7 $M_\odot$ yr$^{-1}$ 
\citep[according to prescriptions of ][]{K99}
and $L$(H$\beta$) $<$ 3$\times$10$^{40}$ erg s$^{-1}$ 
(small blue open circles). Fig. \ref{fig1}b shows the histogram of colour 
distributions for both samples. It is seen from Figs. \ref{fig1}a,b that a 
major fraction of galaxies has $W1-W2$ $\leq$ 0.5 mag, implying that emission 
in these galaxies is dominated by stars and ionised gas. 
On the other hand, there are 65 galaxies with $W1-W2$ $\sim$ 1 -- 2 mag where 
the contribution of hot dust emission is significant. Most of these galaxies
are characterised by high $SFR$s $\geq$ 0.7 $M_\odot$ yr$^{-1}$ and 
can therefore be classified as LCGs \citep{I11} or ``active'' galaxies following the 
nomenclature of \citet{HH04}. No colour dependence on metallicity is present, 
at variance with the conclusion made by \citet{G11} on the base of a smaller 
BCD sample.

In Fig. \ref{fig1}c,d the same sample is split into sources with low and high 
H$\beta$ equivalent widths EW(H$\beta$), which is a measure of star formation 
burst age (high values relate to very recent bursts). Clearly, red 
$W1-W2$ colours are observed mainly in galaxies with
very young ($<$ 3 - 4 Myr) starbursts. 

Figures \ref{fig2}a,b present the dependence of $W1-W2$ colours on the redshift
$z$ and the ($W1-W2$) - ($W2-W3$) colour-colour diagram, respectively, 
for the same sample as in Fig. \ref{fig1} split into two parts
with $L$(H$\beta$) $\geq$ 3$\times$10$^{40}$ erg s$^{-1}$ (red symbols) and
$L$(H$\beta$) $<$ 3$\times$10$^{40}$ erg s$^{-1}$ (blue symbols). No evident
trend of the $W1-W2$ colour on $z$ is present for galaxies with 
$W1-W2$ $\geq$ 1 mag, while for galaxies with bluer $W1-W2$ colours
there is a tendency to be redder with increasing $z$. The colour-colour
diagram (Fig. \ref{fig2}b) shows that galaxies with low $SFR$s (blue symbols) occupy
the region of spiral galaxies, according to \citet{L11}. On the other hand,
galaxies with high $SFR$s (red symbols) are mixed with QSOs,
Seyferts, ultraluminous infrared galaxies (ULIRGs), LINERs and starbursts. 
Physical properties of our galaxies are very different from those of other 
types of galaxies, excluding starbursts. Evidently, the colour-colour diagram
in Fig. \ref{fig2}b fails to separate galaxies of different types.

Coming back to Figs. \ref{fig1}a,c, it can also be seen 
that galaxies with very red colours 
$W1-W2$ $>$ 2 mag are rare. 
\citet{G11} found three galaxies (red filled squares) that are not present 
in our sample, because no SDSS spectra are available for them.
We find only four more galaxies like these
(large red filled circles) out of $\sim$ 5000 galaxies from our SDSS sample 
with the available {\sl WISE} data. 
In the colour-colour diagram our four objects together with
three galaxies from \citet{G11};
(large symbols in Fig. \ref{fig2}b) are located in the region of ULIRGs,
LINERs and obscured AGNs, but their other properties are very different.
General characteristics of the newly identified four galaxies are shown
in Table \ref{tab1}, SDSS images and 
observed spectral energy distributions (SEDs) are shown in 
Figs. \ref{fig3} and \ref{fig4}. All four galaxies are very compact 
($\sim$ 1 - 2\arcsec\ in diameter, corresponding to linear scales 
$\sim$ 3 -- 6 kpc at their redshifts of 0.15 - 0.33) and are almost unresolved.
The SDSS spectra resemble those with
strong emission lines, which is suggestive of active star-formation 
in young bursts with rates $SFR$ $\sim$ 6 - 27 $M_\odot$ yr$^{-1}$. 
Their MIR SEDs (red symbols and solid lines in Fig. \ref{fig4}) 
indicate a 
flux excess similar to that in the ``active'' BCD SBS 0335--052E (red dotted
lines) in contrast to no flux excess in the ``passive'' BCD I Zw 18 (blue 
dashed lines).
Their low stellar masses of $\sim$ 10$^8$ -- 10$^9$ $M_\odot$ 
(Table \ref{tab1}) derived from the
fitting of SDSS spectra are characteristic 
of dwarf galaxies according to \citet{I11}. 

Certainly, 
these galaxies can be classified as ``green pea'' galaxies \citep{C09} or
LCGs \citep{I11}. The characteristics of the newly identified four 
galaxies with very red
$W1-W2$ colours are similar to those in high-redshift LBGs. On the other
hand, oxygen abundances in these galaxies are higher than in the three BCDs 
discussed by \citet{G11}, indicating that there is no apparent dependence
on metallicity. In the sample of 803 LCGs by \citet{I11} there are many
other galaxies with optical characteristics similar to the LCGs in 
Table \ref{tab1} but with more moderate characteristics in the MIR range.
Probably, high H$\beta$ luminosity or young age of the burst is not the only
factor for heating the dust to high temperatures. Other factors may
play a role, such as the morphology and compactness of the galaxy, 
the presence of super-star clusters (SSCs), distribution
of the interstellar gas and dust in the vicinity of young clusters. Therefore,
{\sl Hubble Space Telescope} ({\sl HST}) high-angular resolution imaging,
spectroscopic observations in a wide wavelength range including the
MIR range and
interferometric observations in the H {\sc i} $\lambda$21 cm line and in the
radio-continuum will be useful for a better understanding of the origin of hot dust 
and the determination of its properties.

\section{Summary \label{sum}}

We carried out a search for star-forming Sloan Digital Sky Survey
(SDSS) galaxies with strong emission lines in the 
{{\sl Wide-field Infrared Survey Explorer}} ({{\sl WISE}}) Preliminary Release 
Source Catalogue (PRSC) aiming to find galaxies with hot dust emission
at wavelengths $\lambda$3.4 - 4.6 $\mu$m ($W1$ and $W2$ bands).
We find that $\sim$ 5000 galaxies out of the total sample of $\sim$ 16000 SDSS 
galaxies are present in the PRSC, which covers only $\sim$ 57\% of the sky. 
About 1000 galaxies out of the galaxies detected with 
{\sl WISE} have a sufficiently strong [O {\sc iii}] $\lambda$4363 emission line
in their SDSS spectra to allow a reliable abundance determination.
The comparison of optical and mid-infrared properties for the sample
of galaxies with reliably derived oxygen abundances led us to following 
conclusions:

A major fraction of galaxies has $W1-W2$ of $\sim$ 0.0 - 0.4 mag,
consistent with the colour for the emission from stars and the ionised 
interstellar medium. The
contribution of hot dust emission is small. On the other hand, there are 65
galaxies with redder $W1-W2$ colours, $\sim$ 1 - 2 mag.
Most of these galaxies are luminous compact galaxies (LCGs) with
star-formation rates $SFR$ $>$ 0.7 $M_\odot$ yr$^{-1}$ and high
H$\beta$ equivalent widths EW(H$\beta$) $>$ 50\AA,
implying a very recent starburst, that can efficiently heat interstellar dust
to high temperatures.

Star-forming galaxies with very red $W1-W2$ colours $>$ 2 mag are rare.
In addition to three galaxies previously studied by
\citet{G11}, which are not present in our sample because of the lack of SDSS 
spectra for them,
we find only four galaxies like these from the sample of $\sim$ 5000 galaxies
detected with {\sl WISE}.

\acknowledgements
We thank the referee S. Bianchi for valuable comments.
Y.I.I., N.G.G. and K.J.F. are grateful to the staff of the Max Planck 
Institute for Radioastronomy (MPIfR) for their warm hospitality. 
Y.I.I. and N.G.G. acknowledge a financial support of the MPIfR.
This publication
makes use of data products from the {\sl Wide-field Infrared
Survey Explorer}, which is a joint project of the University of
California, Los Angeles, and the Jet Propulsion Laboratory,
California Institute of Technology, funded by theNational Aeronautics
and Space Administration. Funding for the Sloan
Digital Sky Survey (SDSS) and SDSS-II has been provided by
the Alfred P. Sloan Foundation, the Participating Institutions,
the National Science Foundation, the U.S. Department of
Energy, the National Aeronautics and Space Administration,
the Japanese Monbukagakusho, and the Max Planck Society,
and the Higher Education Funding Council for England.

\end{document}